\documentclass{PoS}

\def\be{\begin{equation}}
\def\ee{\end{equation}}
\def\bea{\begin{eqnarray}}
\def\eea{\end{eqnarray}}

\def\gev{ \hbox{GeV} }

\newcommand{\lsim}{\raise.3ex\hbox{$<$\kern-.75em\lower1ex\hbox{$\sim$}}}
\newcommand{\ima}{{\mbox{Im}\,}}
\newcommand{\rea}{{\mbox{Re}\,}}

\title{Extending the Regge description of hadronic forward scattering from the multi-TeV down to the GeV region}

\ShortTitle{Extending the Regge description of hadronic forward scattering from the multi-TeV down to the GeV region}

\author{\speaker{Jose R. Pelaez}\thanks{Work partially supported by CGICYT Spain, under contracts FPA2005-02327 and BFM2002-01868 and the EURIDICE network contract HPRN-CT-2002-00311, as well as the EU Hadron Physics Project, contract number RII3-CT-2004-506078.}\\
        Departamento de F\'{\i}sica Te\'orica II, Universidad Complutense, 28040, Madrid, Spain\\
        E-mail: \email{jrpelaez@fis.ucm.es}}

\abstract{
We provide a simple Regge parametrization of forward hadronic scattering
from the multi-TeV range down to $\sim1\gev$ above the threshold of each reaction.
 We show how, at these relatively low
energies, mass effects are relevant and should be properly taken into account,
and that the data favor a 
logarithmic growth of the Pomeron contribution that is based on an improved 
unitarity bound at intermediate energies as well as a separate factorization
of the singularities.
Data on both the imaginary  and real parts of amplitudes are
remarkably well described with this parametrization. Here we also show
that the description does not depend on the strategy adopted to include or not 
systematic 
uncertainties in different data sets.}

\FullConference{International Europhysics Conference on High Energy Physics\\
                 July 21st - 27th 2005\\
                 Lisboa, Portugal}

\begin{document}

Recently \cite{Pelaez:2003eh} we extended 
down to $\sim1 \hbox{GeV}$ above threshold 
the Regge description of $(\bar{p}p+pp)$,
$(K^ +p+K^-p)$, $\pi^ \pm N$ and $\pi\pi$ cross sections,
which only involve the 
Pomeron, $f$ (or $P'$) and $\rho$ trajectories.
I present here {\it preliminary results} \cite{inprep} adding also
the $a$ and $\omega$ trajectories and extending the analysis
to $\bar{p}p$, $pp$, $\bar{p}n$, $pn$, $K^\pm p$, $K^\pm n$, $\pi^\pm N$
and $\pi\pi$ total cross sections, and $Im F/ Re F$ ratios for the
$\bar{p}p$, $pp$, $pn$, $\pi^\pm N$ and  $K^\pm p$
forward elastic amplitudes, $F$. 
The data come from the COMPAS group compilation. 
However, the original references did not treat
systematic uncertainties uniformly, and many 
data sets are incompatible within their statistical
errors, and hence cannot be described simultaneously.
Thus, we adopt two fitting strategies:
First, we keep the original uncertainties so that we can easily
compare with 
the PDG \cite{PDG} and the reference works of the COMPETE group \cite{Cudell:2001pn}.
However, this introduces an artificially large $\chi^2/d.o.f.$ and 
a bias toward
those data sets that do not provide systematic uncertainties.
Hence, in our second strategy, we add a
systematic error, but only to those data
without it, of
 $0.5\%$ for $pp$,
$1\%$ for $\bar{p} p$ and $1.5\%$ for other processes.
These additional errors are similar to those given by 
other experiments, and thus all sets are equally weighted. 
To account for different ways of combining
statistical and systematic errors, in the first strategy we have added them
in quadrature and linearly
in the second.
In addition, we use $\sigma^{total}$  data \cite{datapipi} on  $\pi^+\pi^-$, 
$\pi^-\pi^-$,$\pi^+\pi^0$, above $1.42\gev$, plus one data point per
channel reconstructed from phase shift analyses \cite{Pelaez:2003eh} 
at $1.42\gev$. If using $\pi\pi$ low energy information \cite{Pelaez:2003eh,inprep},
the $\rho$ residue and the intercept come out somewhat smaller 
and larger, respectively.

The different Regge trajectories contribute to the amplitudes as follows:
\begin{eqnarray}
F_{p^\pm p}=(P_{NN}+f_{NN}+a_{NN}\mp
\omega_{NN}\mp\rho_{NN})/2, \;
F_{p^\pm n}=(P_{NN}+f_{NN}-a_{NN}\mp
\omega_{NN}\pm\rho_{NN})/2,\nonumber\\
F_{K^\pm p}=(P_{KN}+f_{KN}+a_{KN}\mp
\omega_{KN}\mp\rho_{KN})/2,\;
F_{K^\pm n}=(P_{KN}+f_{KN}-a_{KN}\mp
\omega_{KN}\pm\rho_{KN})/2,\nonumber\\
F_{\pi^\pm p}=(P_{\pi N}+f_{\pi N})/\sqrt{6}\mp\rho_{\pi N}/2,
\quad F_{\pi^\pm\pi^-}= (P_{\pi\pi}+f_{\pi \pi})/3\pm\rho_{\pi \pi}/2,
\quad
F_{\pi^0\pi^-}= (P_{\pi\pi}+f_{\pi \pi})/3, \nonumber
  \label{eq:Fsandpoles}
\end{eqnarray}
where $N=p^\pm,n$ and we use the factorization \cite{factorization}
relations $R_{AB}(\nu)=f^R_A 
f^R_B R(\nu)$, with 
\begin{equation}
  \label{eq:factorization}
R(\nu)= \beta_R 
\left(\frac{1+\tau e^{-i\pi\alpha}}{\sin\pi\alpha}\right)
\nu^{\alpha_R},\quad \hbox{for}\; R=\rho,f,a,\omega,
%\left(\frac{\nu}{\nu_0}\right)^{\alpha_R},
%\; \nu_0=1\hbox{GeV}^2.
\end{equation}
where $\tau$ is the signature of the trajectory.
Masses are correctly taken into account by using the Regge variable
{$\nu=(s-u)/2$}, which, for forward scattering, is
 $\nu=s-m_a^2-m_b^2>s-s_{th}$. Following the QCD version of Regge theory, 
and the recent analysis \cite{Cudell:2001pn}, we assume $\alpha_a=\alpha_f$ and 
$\alpha_\omega=\alpha_\rho$.
We set $f^R_\pi=1$,
for $R=P,f,\rho$ and $\beta_R=1$ for $R=a,\omega$, since they are redundant. 
%the comparison with previous works \cite{Pelaez:2003eh}.

For the Pomeron, we propose the use of a ``constant plus logarithm'' law, i.e.,
\begin{equation}
  \label{eq:pomeron}
P_{AB}=C_{AB}+L_{AB}, \quad\ima P(\nu)=\nu \,\left(\beta_P+A\log^2
\left[
\frac{\nu-\nu_{th}}{\nu_1\log^{7/2}(\nu/\nu_2)}
\right]
\right),
%\left(\frac{\nu}
%{\nu_0}\right)
\end{equation}
where $\nu_{th}$ is the right-cut branch point of each amplitude.
This law follows the improved
unitarity bound in \cite{Yndurain:1973rx} that
grows faster than $s\log s$
but slower at intermediate energies than the $s\log^2 s$ 
Froissart bound, which is 
recovered at very high $s$. 
The generalized ``factorization theorem'' \cite{Cudell:2002ej} 
requires 
singularities to factorize separately, and as a first approximation we thus use separated
$f_A^C$ and $f_A^L$.

Finally, the $\rea F$ 
are obtained from dispersive representations, and total cross sections from:
\begin{eqnarray}
\sigma_{ab}=4\pi^ 2 \hbox{Im}\, F_{a+b\rightarrow a+b}(s,0)
/\lambda^{1/2}(s,m_a^2,m_b^2), \quad
\lambda(s,m_a^2,m_b^2)=s^2+(m_a^2-m_b^2)^2-2s(m_a^2+m_b^2).
\nonumber
\label{eq:totcrossamp}
\end{eqnarray}
Let us remark that $\lambda$
is usually approximated by $s^2$, although very recently \cite{Cudell:2003ci}
a slight improvement in $\chi^2/d.o.f.$ has been reported
using $\lambda$, instead of $s^ 2$, down to $\sqrt{s}=5\,$GeV. 
Note that, when $E_{kin}\simeq 1\,$GeV, as in our case here, 
the use of $s^2$, instead of $\lambda$,
yields a 30\% overestimation for $NN$.

Different sets of parameters from fits to data
using the two strategies described above are shown in Table 1, 
including
the uncertainty obtained from the $\chi^2/d.o.f.$ minimization (using
MINUIT), for strategy 2, whose $\chi^2/d.o.f.=0.85$ for 1186 data points.
Due to the strong correlations,
amplitudes should be calculated using with parameters within the same set.
However, systematic 
errors for {\it a single parameters} can be estimated from the
difference between strategies.  
\vspace*{.2cm}

\hspace*{-.5cm}
%\vspace*{.2cm}
  {\centerline
  {\footnotesize
\begin{tabular}{|c|c|c||c|}
\hline
%& Our fit & Our fit&Minuit\\
& strategy 2& strategy 1&Minuit\\
$E_{kin}^{min}$& 1-1.3 GeV& 1-1.3 GeV&errors\\ \hline
$\beta_P$ &0.746&0.937&0.003\\
$f_N^P$ &1.792&1.705&0.007\\
$f_K^P$ &0.731&0.714&0.004\\
$A$ &0.043&0.050&0.001\\
$\nu_1$ &0.0005&0.001&0.0001\\
$\nu_2$ &0.676&0.633&0.001\\
$f_N^{log}$ &1.02&0.993&0.001\\
$f_K^{log}$ &0.723&0.733&0.012\\ 
$\beta_f$ &1.70&1.77&0.014\\
$f_N^f$ &1.78&1.75&0.01\\ \hline
\multicolumn{3}{c}{ }
\end{tabular}
\begin{tabular}{|c|c|c||c|}
\hline
%& Our fit & Our fit&Minuit\\
& strategy 2& strategy 1&Minuit \\
$E_{kin}^{min}$& 1-1.3 GeV& 1-1.3 GeV&errors\\ \hline
$f_K^f$ &0.30&0.32&0.01\\
$\alpha_f$ &0.646&0.640&0.002\\
$f_N^a$ &-0.24&0.25&0.04\\
$f_K^a$ &-0.55&0.5&0.1\\ 
$\beta_\rho$ &1.28&1.34&0.11\\
$f_N^\rho$ &0.51&0.46&0.04\\
$f_K^\rho$ &0.49&0.54&0.04\\
$\alpha_\rho$ &0.464&0.464&0.003\\
$f_N^\omega$ &1.97&1.98&0.015\\
$f_K^\omega$ &0.66&0.65&0.01\\
\hline
$\sigma_{LHC}$&109 mb&110 mb&1mb\\
\hline
\end{tabular}
}
}
{\footnotesize {\bf Table 1.} Fit parameters with different strategies.
The Minuit errors are {\it just statistical}, and nominal, since the parameters are strongly correlated
and can only be used with the central values of {\it strategy 2}. 
} 
\label{elesln}
%\end{table}

In Table 2 we show how the $\chi^2/d.o.f.$ deteriorates if 
we do not implement one of the following items: i)  
 using $\nu$ instead of $s$, ii) the logarithmic growth in Eq.(\ref{eq:pomeron}),
iii) the separate factorization.

{\footnotesize
\begin{minipage}{\textwidth}
\vspace*{.2cm}
\begin{center}
\begin{tabular}{|c|c|c|c|c|}
\hline
$E_{kin}^{min}$ (GeV)&1-1.3& 1.5 & 2 & 3 \\ \hline
$\#$ data points& 1186 & 1002 &  895 & 768 \\ \hline
\hline Parametrization 
&
\multicolumn{4}{|c|}{$\chi^2/d.o.f.$ for strategy 2 / 1}  \\ \hline
Ours &0.85/1.56&0.63/1.14&0.57/1.05&0.52/0.95\\ \hline
$\nu_1\equiv0.01$ GeV$^2$ &0.85/1.57&0.63/1.26&0.58/1.06&0.52/0.97\\
\hline\hline
powers of $s^\alpha$ &1.58/2.87&1.16/2.11&0.99/1.80& 0.78/1.42\\\hline \hline 
\multicolumn{5}{|c|}{Pomeron logarithmic term}  \\
\hline 
$\nu\, log (\nu)$ &1.01/1.83&0.69/1.26&0.59/1.09 &0.52/0.97\\
\hline
$\nu\, log (\nu-\nu_{th})$ &1.03/1.83&0.69/1.27&0.59/1.12 &0.52/0.98\\
\hline $\nu\, log^2 (\nu)$ &0.97/1.79&0.68/1.24&0.59/1.10 &0.52/0.95\\
\hline
$\nu\, log^2 (\nu-\nu_{th})$ &0.91/1.68&0.65/1.18&0.58/1.06 &0.52/0.95\\
\hline \hline 
\multicolumn{5}{|c|}{Factorization of Pomeron logarithms}  \\
\hline 
$f_a^L\equiv1$ (as PDG) &0.92/1.70&0.66/1.23&0.59/1.10 &0.54/1.01\\
\hline 
$f_a^C=f_a^L$ &0.89/1.67&0.64/1.41&0.60/1.14&0.58/1.02\\
\hline
\end{tabular}

\vspace{.2cm}
{\footnotesize {\bf Table 2.} $\chi^2/d.o.f.$ for several
  $E_{kin}^{min}$ and different modifications of our parametrization.
}
\vspace*{.2cm}
\end{center}
\end{minipage}
}

In \cite{inprep}, we have already shown in plots that our parametrization provides a remarkable
description of total $NN$, $\pi\pi$, $\pi^ \pm N$,$K^\pm p$ and $K^\pm n$ 
cross sections and $Re F/Im F$. The simple parametrization reported here was shown to
describe remarkably well 20 observables
extending from several TeV down to $\sim 1\gev$ above the threshold
of each reaction. We will not repeat the plots here and instead we will show
that, although the central values of each parameter for strategy 2
could be beyond
one standard deviation from those of strategy 2, when considering the
complete parameter sets, the results overlap. Indeed, 
we show in Figure 1 the curves obtained from our parametrization from strategy 2
(the continuous line and  gray bands covering
 its nominal uncertainties), versus the results for strategy 1 (the dashed line). Although
we have chosen the plot where they deviate most, both curves are almost indistinguishable.

Further details will be given in
a forthcoming publication \cite{inprep}.
We hope that, apart from establishing the logarithmic
growth of the Pomeron, our parametrization could be easily used for 
dispersive studies in hadronic physics that involve integrals
from the resonance region to infinity.

\begin{figure}[htbp]
  \centering
  \includegraphics[width=.47\textwidth,angle=-90]{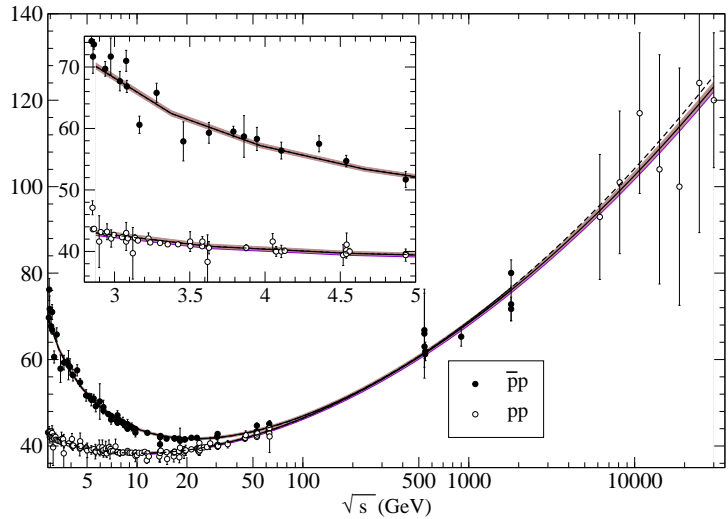}
  \caption{\footnotesize Total $pp$ and $\bar pp$
cross sections. Results from our parametrization down
to 1 GeV above threshold both for strategy 2 (continuous line)
and strategy 1 (dashed line). Note that both curves basically overlap
over the bands that cover the nominal uncertainties
in the parameters. For other processes the agreement between strategies is even better.
}
  \label{fig:2}
\end{figure}

\end{document}